\documentstyle[sprocl,epsf]{article}

\def\NCA{{\em Nuovo Cimento }}

\def\NPB{{\em Nucl. Phys.} B}
\def\PLB{{\em Phys. Lett.}  B}
\def\PRL{{\em Phys. Rev. Lett.}}
\def\PRD{{\em Phys. Rev.} D}
\def\ZPC{{\em Z. Phys.} C}

\newcommand{\nn}{\nonumber}
\newcommand{\be}{\begin{equation}}
\newcommand{\ee}{\end{equation}}
\newcommand{\bea}{\begin{eqnarray}}
\newcommand{\eea}{\end{eqnarray}}

\def\bfnabla{\mbox{\boldmath $\nabla$}}

\def\bfsigma{\mbox{\boldmath $\sigma$}}
\def\bfPi{\mbox{\boldmath $\Pi$}}
\def\lQ{\Lambda_{\rm QCD}}

\def\als{\alpha_{\rm s}}
\def\siml{{\ \lower-1.2pt\vbox{\hbox{\rlap{$<$}\lower6pt\vbox{\hbox{$\sim$}}}}\ }}

\begin{document}

\title{pNRQCD: REVIEW OF SELECTED RESULTS}
 
\author{ANTONIO VAIRO\footnote{Invited talk presented at "4th International 
Conference on Quark Confinement and the Hadron Spectrum", Vienna, 3--8 July 
2000.} } 
\address{Institut f\"ur Theoretische Physik, Universit\"at Heidelberg\\ 
Philosophenweg 16, D-69120 Heidelberg, Germany \\ E-mail: a.vairo@thphys.uni-heidelberg.de}

\maketitle\abstracts{I review and discuss a selected sample of recent results in pNRQCD.}

\section{Introduction}
Non-relativistic bound-state systems are characterized by, at least, three widely separated scales: 
the mass $m$ of the particle, the (soft) scale associated to its relative momentum 
$\sim mv$, $ v \ll 1$, and the (ultrasoft) scale associated to its kinetic energy $\sim mv^2$. 
In QED and in the perturbative regime of QCD the velocity $v$ of the particle in the bound state  
may be identified with the coupling constant. Moreover, the inverse of the size of the system 
is also of order $m v$ and the binding energy of order $m v^2$. Indeed, a systematic treatment 
of non-relativistic bound-state systems in the framework of effective field theories (EFT), which takes 
full advantage of the above energy scale hierarchy, was initiated in QED \cite{NRQED} and in 
more recent years remarkable progress has been achieved in the analysis of $t\bar{t}$ threshold 
production \cite{topnnlo}. 

\begin{figure}[th]
\makebox[0.0truecm]{\phantom b}
\put(70,0){\epsfxsize=6.8truecm \epsffile{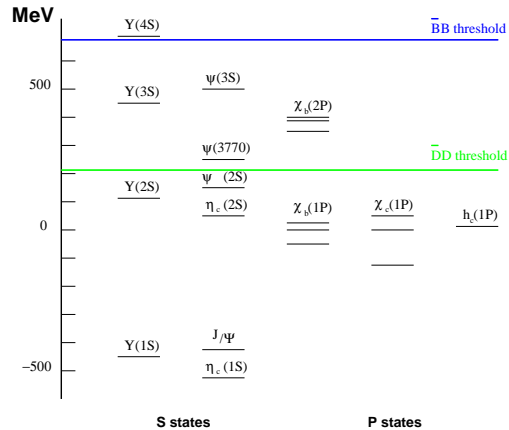}}
\caption{\it The spectra of $b\bar{b}$ and $c\bar{c}$ quarkonia normalized with respect 
 to the spin average of the $\chi_b(1P)$ and $\chi_c(1P)$ states respectively.}
\vspace{-3mm}
\label{masscale}
\end{figure}

For systems made of $b$ and $c$ quarks (I will denote them generically as heavy quarkonia: 
$\psi$, $\Upsilon$, $B_c$, ...) non-perturbative contributions may be relevant. By comparing the energy 
level spacings of these systems (see Fig. \ref{masscale}) with the heavy-quark masses 
(e.g. $m_b \simeq 5$ GeV and $m_c \simeq 1.6$ GeV) we can still argue that the data are consistent 
with a kinetic energy of the bound quark much smaller than the heavy-quark mass and, therefore,  
with a non-relativistic (NR) description of the heavy-quark--antiquark system. 
However, in dependence of the specific system, the scale of non-perturbative physics,  
$\lQ$, may turn out to be close to some of its dynamical scales.
The physical picture, which then arises, may be quite different from the perturbative 
situation. What remains guaranteed, also for heavy quarkonia, is that $m \gg \lQ$ and that at least the mass 
scale can be treated perturbatively, i.e. integrated out from QCD order by order in the coupling constant. 
The resulting EFT is called NRQCD \cite{NRQCD}. 

A lot of effort has been put over the last two decades in order to find the relevant 
operators, which parameterize the non-perturbative heavy-quark--antiquark interaction,  
once the mass scale has been integrated out. In some classical works \cite{wilson,Brown,spin,BMP} 
these operators were identified with Wilson loop operators. At the same time, however, the relevance 
of less extended non-perturbative objects was pointed out in \cite{VL,gro82,Balitsky} for situations 
where the scale of non-perturbative physics is of the order $m v^2$ or smaller. A first non-perturbative 
derivation of some heavy quarkonia potentials in the framework of NRQCD was done in \cite{chen}. 
While a full systematic study of the heavy-quark--antiquark systems in an EFT framework, 
which incorporates all the possible dynamical situations (at least in pure gluodynamics) and 
factorizes the relevant non-perturbative operators, has been recently completed in \cite{long,m1,m2}.

\begin{figure}[\protect t]
\makebox[0.0truecm]{\phantom b}
\put(65,0){\epsfxsize=7.5truecm \epsffile{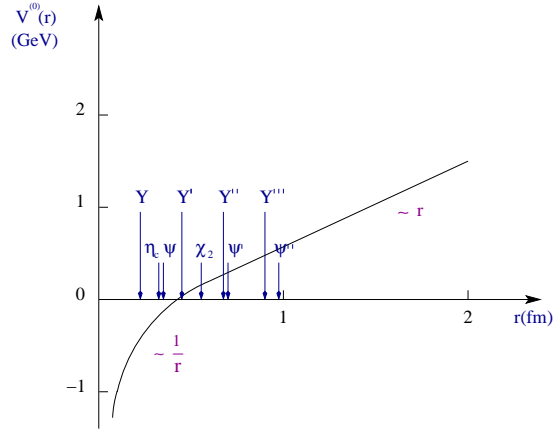}}
\caption{\it The size of some heavy quarkonia, as calculated in \protect \cite{gois},  
is shown with respect to the quark-antiquark static potential.}
\vspace{-5mm}
\label{plpot}
\end{figure}

In the following I will discuss the EFT that may be constructed from NRQCD by integrating out 
the scale of the momentum transfer, assumed to be the next relevant scale of the system.  
I shall call the obtained EFT, potential NRQCD (pNRQCD) \cite{pNRQCD}. 
In Sec. \ref{sec1} I will consider the situation where this scale is much bigger than $\lQ$,
(more specifically I will consider $mv^2$ not smaller than $\lQ$). To this situation belong QED bound 
states (in the appropriate gauge-group limit) and what would be $t\bar{t}$ bound states. 
In particular, I will review the $\alpha^5 \ln \alpha$ calculation of the quarkonium spectrum 
and some of its implication for the $e^+e^- \rightarrow t\bar{t}$ cross section.
It is not a priori clear to which heavy-quarkonium states these results apply. 
As a guideline we may take the results of \cite{gois} plotted in Fig. \ref{plpot}. 
Eventually the internal consistency of the EFT and the comparison with the experimental 
data will provide a way to discriminate among the different situations. 
The quarkonium ground-state radii, in particular for the $\Upsilon(1S)$, appear to fall in a region 
where the potential is characterized by a Coulomb-type behaviour. This suggests that a perturbative 
treatment at the momentum-transfer scale may be correct. Instead, heavy-quarkonium resonances higher 
than the ground state fall in a region where the potential is no longer of the Coulomb-type. 
This seems to indicate that a perturbative treatment of the momentum transfer scale is not allowed for them.
In Sec. \ref{sec2} I will consider this last situation.

\section{Quarkonium at the NNNLO}
\label{sec1}
In this section I shall discuss heavy quarkonium in the dynamical situation where $m v^2$ 
is not smaller than $\lQ$. This means that at the matching scale to pNRQCD, $m v > \mu > mv^2$, 
I can still assume that (ultrasoft) gluons and quark-antiquark states in color-singlet and color-octet 
configuration exist. What would be toponium in $t\bar{t}$ threshold production 
and (likely) heavy-quarkonium ground states fall in this situation. The aim is to set up the framework 
for an eventual full NNNLO calculation of the heavy quarkonium masses as well as 
the $e^+e^- \rightarrow t\bar{t}$ cross section. More explicitly I will give the leading log 
contributions to the NNNLO. 

The pNRQCD Lagrangian in the situation $\lQ \siml mv^2$, considering only the terms relevant 
to the analysis of the leading-log corrections at the NNNLO of the singlet, reads as follows:
\begin{eqnarray}                         
&& \hspace{-6mm} {\cal L}_{\rm pNRQCD} =
{\rm Tr} \,\Biggl\{ {\rm S}^\dagger \left( i\partial_0 - {{\bf p}^2\over m} +{{\bf p}^4\over 4m^3}
- V^{(0)} - {V^{(1)} \over m}- {V^{(2)} \over m^2}\right) {\rm S} \nn \\
&& \hspace{-2mm} + {\rm O}^\dagger \left( iD_0 - {{\bf p}^2\over m} - V^{(0)}_o \right) {\rm O} \Biggr\}
\qquad - {1\over 4} G_{\mu \nu}^{a} G^{\mu \nu \, a} \nn\\
&&  \hspace{-2mm} + g V_A {\rm Tr} \left\{  {\rm O}^\dagger {\bf r} \cdot {\bf E} \,{\rm S}
+ {\rm S}^\dagger {\bf r} \cdot {\bf E} \,{\rm O} \right\} 
+ g {V_B  \over 2} {\rm Tr} \left\{  {\rm O}^\dagger {\bf r} \cdot {\bf E} \, {\rm O} 
+ {\rm O}^\dagger {\rm O} {\bf r} \cdot {\bf E}  \right\}\!,   
\label{pnrqcdph}
\end{eqnarray}
where ${\bf r}$ is the relative coordinate, ${\bf p} = -i \bfnabla_{{\bf r}}$, and 
S ($= S \, { 1\!{\rm l}_c/\sqrt{N_c}}$) and O are the singlet and octet field, respectively.  
All the gauge fields in Eq. (\ref{pnrqcdph}) are functions of the centre-of-mass coordinate 
and the time $t$ only. 

\begin{figure}
\makebox[0.0cm]{\phantom b}
\put(15,10){\epsfxsize=4.5truecm \epsfbox{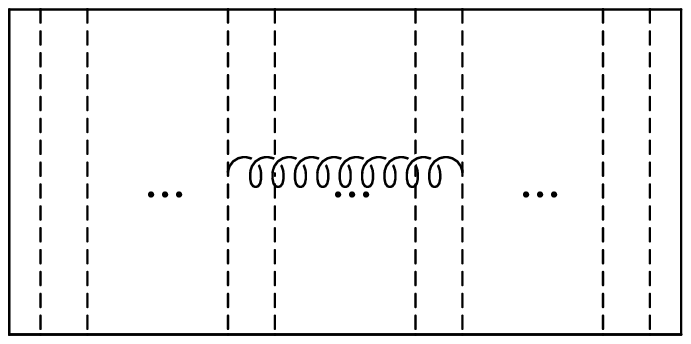}}
\put(153,37){\epsfxsize=1.0truecm \epsfbox{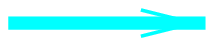}}
\put(193,30){\epsfxsize=4.5truecm \epsfbox{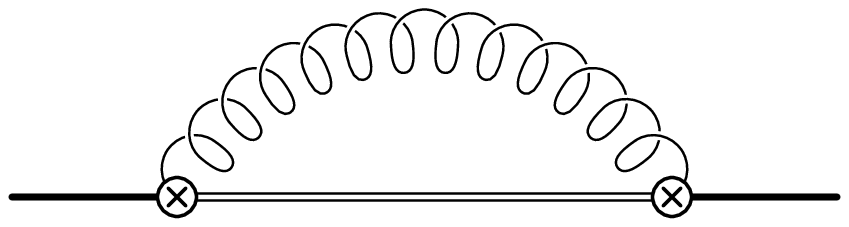}}
\vspace{-2mm}
\caption{The Feynman graphs of the static Wilson loop (left) contributing to the three-loop leading logs 
of the static matching potential and the corresponding graph in pNRQCD (right). 
The double line indicates the octet propagator.}
\vspace{-3mm}
\label{US}
\end{figure}

The functions $V$ are the matching coefficients of pNRQCD. Here I am interested only in the singlet 
sector at $\als^5\ln \als$ accuracy. For this purpose $V^{(0)}$ can be obtained from matching NRQCD 
to pNRQCD at $O(1/m^0)$ exactly at the two-loop level \cite{1loop,twoloop} and with leading-log 
accuracy at the three-loop level \cite{long} (see Fig. \ref{US}). The result reads
\bea
&&V^{(0)} =-C_F{\alpha_{\rm s}(r)\over r}
\left\{1+\left(a_1+ 2 {\gamma_E \beta_0}\right) {\alpha_{\rm s}(r) \over 4\pi}\right.
\nonumber\\
&&\quad +\left[\gamma_E\left(4 a_1\beta_0+ 2{\beta_1}\right)+\left( {\pi^2 \over 3}
+4 \gamma_E^2\right) 
{\beta_0^2}+a_2\right] {\alpha_{\rm s}^2(r) \over 16\,\pi^2}
\left. + {C_A^3 \over 12}{\alpha_{\rm s}^3 \over \pi} \ln{\mu r}\right\},
\label{newpot0}
\eea 
where $\beta_n$ are the coefficients of the beta function and the values of $a_1$ and $a_2$ 
can be found in \cite{twoloop}. For the calculation of the matching potential $V^{(1)}$ 
and $V^{(2)}$ we need to perform the matching exactly at one-loop level \cite{Duncan} and with  
leading-log accuracy at two-loop level \cite{nnnlo} and exactly at tree level \cite{Gupta}  
and with leading-log accuracy at one loop level \cite{nnnlo} respectively. 
(The leading logs may be extracted by considering the ultraviolet divergences of the 
pNRQCD diagram in Fig. \ref{US}, when evaluated on the full octet propagator.) 
The result reads 
($S_{12}({\hat {\bf r}}) = 3 {\hat {\bf r}}\cdot \bfsigma_1 \,{\hat {\bf r}}\cdot \bfsigma_2 
- \bfsigma_1\cdot \bfsigma_2$, ${\bf S}_j = \bfsigma_j/2$)
\begin{eqnarray}
V^{(1)} &=& -C_F C_A {\alpha_{\rm s}^2(r)\over 2 r^2}
\left\{1+{2 \over 3}(4C_F+2C_A){\als \over \pi}\ln{\mu r} \right\},
\label{Vs1} \\
V^{(2)} &=& \left\{{\bf p}^2,V_{{\bf p}^2}^{(2)}(r)\right\}
+ {V_{{\bf L}^2}^{(2)}(r)\over r^2}{\bf L}^2 + V_r^{(2)}(r) \nn \\ 
&& + V_{LS}^{(2)}(r){\bf L}\cdot({\bf S}_1+{\bf S}_2)
+ V_{S^2}^{(2)}(r) {\bf S}_1\cdot{\bf S}_2 + V_{{S}_{12}}^{(2)}(r){\bf S}_{12}({\hat {\bf r}}), 
\label{VVs2} \\
V_{{\bf p}^2}^{(2)}(r) &=& -{C_F\alpha_{\rm s}(r) \over 2 r} 
\left\{1+{4 \over 3}C_A{\als \over \pi} \ln{\mu r} \right\},
\label{Vs2}\\ 
V_{{\bf L}^2}^{(2)}(r) &=& {C_F \alpha_{\rm s}(r) \over 2 r},
\label{Vs22}\\ 
V_r^{(2)}(r) &=& 3\delta^{(3)}({\bf r}) \pi C_F 
\alpha_{\rm s}(r)\left\{1+ {1\over 9}{\als \over \pi}\left( 2 C_F  
+ {13 C_A\over 2}\right)\ln{m r} \right. \nn\\
& & \left. 
+ {16\over 9}{\als \over \pi}\left({C_A\over 2} - C_F\right)\ln{\mu r}\right\},
\label{Vd2}\\ 
V_{LS}^{(2)}(r) &=& {3 C_F \als(r)\over 2 r^3} 
\left\{1-{2 C_A \over 3}{\als \over \pi}\ln{m r} \right\}, 
\label{VLs2}\\ 
V_{S^2}^{(2)}(r) &=& {8\over 3}  \delta^{(3)}({\bf r}) \pi C_F 
\als(r)\left\{1-{7 C_A \over 4}{\als \over \pi}\ln{m r} \right\} ,
\label{Vss2}\\ 
V_{{S}_{12}}^{(2)}(r) &=& {C_F \als(r) \over 4 r^3}
\left\{1-C_A{\als \over \pi}\ln{m r} \right\}.
\label{Vsten2}
\end{eqnarray}

The quarkonium spectrum at leading log accuracy of the NNNLO, in the situation $\lQ  \siml  m v^2$, 
is given by \cite{nnnlo,KP}
\begin{eqnarray}
E_{n,l,j}  &=&  \langle n,l\vert  V^{(0)} + {V^{(1)} \over  m}  
+ {V^{(2)} \over  m^2 }  \vert n,l \rangle \nn \\
& & \!\!\!\!\!\!\!  -i{g^2 \over 3 N_c}T_F  \int_0^\infty \!\! dt \,
\langle n,l |  {\bf r} e^{it( E_n - h_o)} {\bf r}  | n,l \rangle  
\langle {\bf E}^a (t) \phi(t,0)^{\rm adj}_{ab} {\bf E}^b (0) \rangle ,
\label{nnnlospectrum} 
\end{eqnarray}
where $E_n = -  m  \displaystyle{C_F^2 \alpha_{\rm s}^2  \over 4  n^2}$, 
$h_o = {{\bf p}^2/m } + V_o^{(0)}$ and  the states $|n, l \rangle$ are the eigenstates 
of the Hamiltonian ${{\bf p}^2/ m } + V^{(0)}$.  
The $ \mu$ dependence of the first line of Eq. (\ref{nnnlospectrum}) cancels against the ultrasoft 
contributions of the second line, which corresponds to the pNRQCD diagram of Fig. \ref{US} when 
evaluated on the full octet propagator $i/(E-h_o)$.

\begin{figure}[\protect h]
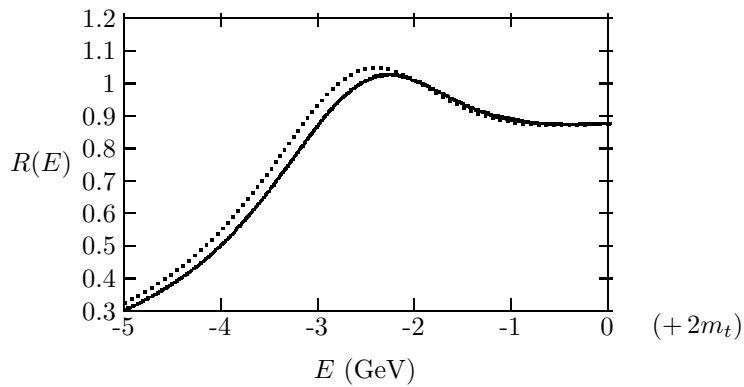

\makebox[0cm]{\phantom b}
\caption{$R(E)$ versus $E$ for the parameter choice $m_t =175$ GeV, $\Gamma_t =1.43$ GeV and 
$\als(M_Z) = 0.118$. The plot is taken from \protect \cite{KP2}. The dotted line corresponds 
to the NNLO calculation, the solid line includes the leading log NNNLO.}
\label{topcros}
\end{figure}

In the situation where $\lQ \ll m v^2$  the correlator 
$\langle {\bf E}^a (t) \phi(t,0)^{\rm adj}_{ab} {\bf E}^b (0) \rangle$
can be calculated perturbatively (a one-loop calculation of it is in \cite{eeper}).
An explicit expression of Eq. (\ref{nnnlospectrum}) at order $\als^5 \ln \als$ is given in \cite{nnnlo}. 
In \cite{KP2} also the ultrasoft corrections to the wave-functions in the origin have been calculated:
\be
\delta \psi^2(0)_{n,0,s} = 
-{m^3 C_F^4 \als^6 \over 8 \pi n^3}
\left\{{3\over2}C_F^2+\left[{41\over12}-{7\over12}s(s+1)\right]C_FC_A  
+ {2\over 3}C_A^2\right\}\ln^2{1\over\als}.
\ee
From these the leading-log correction to the NNNLO of the $e^+e^- \rightarrow t\bar{t}$ 
cross section has been calculated. In Fig. \ref{topcros} $\,R(E) = 
{\sigma(e^+e^- \rightarrow t\bar{t}) / \sigma(e^+e^- \rightarrow \mu^+ \mu^-)}$
is shown at NNLO and with the leading log correction included. 
The aim of such an analysis, once the complete NNNLO will be calculated, 
is to reach  a 50 MeV sensitivity on the top quark mass from the  $t$-$\bar t$ cross-section
near threshold to be measured at a Next Linear Collider \cite{topnnlo}. \vspace{0mm}\\

\begin{center}
% GNUPLOT: LaTeX picture
%\setlength{\unitlength}{0.240900pt}
\setlength{\unitlength}{0.140900pt}
\ifx\plotpoint\undefined\newsavebox{\plotpoint}\fi
\begin{picture}(1500,990)(0,0)
\font\gnuplot=cmr10 at 10pt
\gnuplot
\sbox{\plotpoint}{\rule[-0.200pt]{0.400pt}{0.400pt}}%
\put(181.0,163.0){\rule[-0.200pt]{4.818pt}{0.400pt}}
\put(161,163){\makebox(0,0)[r]{0.3}}
\put(1460.0,163.0){\rule[-0.200pt]{4.818pt}{0.400pt}}
\put(181.0,250.0){\rule[-0.200pt]{4.818pt}{0.400pt}}
\put(161,250){\makebox(0,0)[r]{0.4}}
\put(1460.0,250.0){\rule[-0.200pt]{4.818pt}{0.400pt}}
\put(181.0,338.0){\rule[-0.200pt]{4.818pt}{0.400pt}}
\put(161,338){\makebox(0,0)[r]{0.5}}
\put(1460.0,338.0){\rule[-0.200pt]{4.818pt}{0.400pt}}
\put(181.0,425.0){\rule[-0.200pt]{4.818pt}{0.400pt}}
\put(161,425){\makebox(0,0)[r]{0.6}}
\put(1460.0,425.0){\rule[-0.200pt]{4.818pt}{0.400pt}}
\put(181.0,512.0){\rule[-0.200pt]{4.818pt}{0.400pt}}
\put(161,512){\makebox(0,0)[r]{0.7}}
\put(1460.0,512.0){\rule[-0.200pt]{4.818pt}{0.400pt}}
\put(181.0,600.0){\rule[-0.200pt]{4.818pt}{0.400pt}}
\put(161,600){\makebox(0,0)[r]{0.8}}
\put(1460.0,600.0){\rule[-0.200pt]{4.818pt}{0.400pt}}
\put(181.0,687.0){\rule[-0.200pt]{4.818pt}{0.400pt}}
\put(161,687){\makebox(0,0)[r]{0.9}}
\put(1460.0,687.0){\rule[-0.200pt]{4.818pt}{0.400pt}}
\put(181.0,774.0){\rule[-0.200pt]{4.818pt}{0.400pt}}
\put(161,774){\makebox(0,0)[r]{1}}
\put(1460.0,774.0){\rule[-0.200pt]{4.818pt}{0.400pt}}
\put(181.0,862.0){\rule[-0.200pt]{4.818pt}{0.400pt}}
\put(161,862){\makebox(0,0)[r]{1.1}}
\put(1460.0,862.0){\rule[-0.200pt]{4.818pt}{0.400pt}}
\put(181.0,949.0){\rule[-0.200pt]{4.818pt}{0.400pt}}
\put(161,949){\makebox(0,0)[r]{1.2}}
\put(1460.0,949.0){\rule[-0.200pt]{4.818pt}{0.400pt}}
\put(181.0,163.0){\rule[-0.200pt]{0.400pt}{4.818pt}}
\put(181,122){\makebox(0,0){-5}}
\put(181.0,929.0){\rule[-0.200pt]{0.400pt}{4.818pt}}
\put(441.0,163.0){\rule[-0.200pt]{0.400pt}{4.818pt}}
\put(441,122){\makebox(0,0){-4}}
\put(441.0,929.0){\rule[-0.200pt]{0.400pt}{4.818pt}}
\put(701.0,163.0){\rule[-0.200pt]{0.400pt}{4.818pt}}
\put(701,122){\makebox(0,0){-3}}
\put(701.0,929.0){\rule[-0.200pt]{0.400pt}{4.818pt}}
\put(960.0,163.0){\rule[-0.200pt]{0.400pt}{4.818pt}}
\put(960,122){\makebox(0,0){-2}}
\put(960.0,929.0){\rule[-0.200pt]{0.400pt}{4.818pt}}
\put(1220.0,163.0){\rule[-0.200pt]{0.400pt}{4.818pt}}
\put(1220,122){\makebox(0,0){-1}}
\put(1220.0,929.0){\rule[-0.200pt]{0.400pt}{4.818pt}}
\put(1480.0,163.0){\rule[-0.200pt]{0.400pt}{4.818pt}}
\put(1480,122){\makebox(0,0){0}}
\put(1480.0,929.0){\rule[-0.200pt]{0.400pt}{4.818pt}}
%\put(181.0,163.0){\rule[-0.200pt]{312.929pt}{0.400pt}}%%
%\put(1480.0,163.0){\rule[-0.200pt]{0.400pt}{189.347pt}}%%
%\put(181.0,949.0){\rule[-0.200pt]{312.929pt}{0.400pt}}%%
%\put(181.0,163.0){\rule[-0.200pt]{0.400pt}{189.347pt}}%%
\put(181.0,163.0){\rule[-0.200pt]{180.929pt}{0.400pt}}%%
\put(1480.0,163.0){\rule[-0.200pt]{0.400pt}{110.347pt}}%%
\put(181.0,949.0){\rule[-0.200pt]{180.929pt}{0.400pt}}%%
\put(181.0,163.0){\rule[-0.200pt]{0.400pt}{110.347pt}}%%
\put(-40,556){\makebox(0,0){$R(E)$}}
\put(830,1){\makebox(0,0){$E$ (GeV)}}
\sbox{\plotpoint}{\rule[-0.500pt]{1.000pt}{1.000pt}}%
\put(181,185){\usebox{\plotpoint}}
\put(181.00,185.00){\usebox{\plotpoint}}
\put(199.27,194.84){\usebox{\plotpoint}}
\put(217.55,204.68){\usebox{\plotpoint}}
\put(235.73,214.68){\usebox{\plotpoint}}
\put(253.41,225.56){\usebox{\plotpoint}}
\put(271.08,236.44){\usebox{\plotpoint}}
\put(288.18,248.20){\usebox{\plotpoint}}
\put(304.99,260.37){\usebox{\plotpoint}}
\put(321.83,272.50){\usebox{\plotpoint}}
\put(338.31,285.10){\usebox{\plotpoint}}
\put(354.15,298.51){\usebox{\plotpoint}}
\put(369.99,311.92){\usebox{\plotpoint}}
\put(385.47,325.74){\usebox{\plotpoint}}
\put(400.72,339.82){\usebox{\plotpoint}}
\put(415.45,354.45){\usebox{\plotpoint}}
\put(430.04,369.20){\usebox{\plotpoint}}
\put(444.17,384.41){\usebox{\plotpoint}}
\put(458.29,399.62){\usebox{\plotpoint}}
\put(472.21,415.01){\usebox{\plotpoint}}
\put(485.59,430.88){\usebox{\plotpoint}}
\put(498.68,446.98){\usebox{\plotpoint}}
\put(511.76,463.09){\usebox{\plotpoint}}
\put(524.64,479.37){\usebox{\plotpoint}}
\put(537.24,495.86){\usebox{\plotpoint}}
\put(549.68,512.48){\usebox{\plotpoint}}
\put(561.97,529.20){\usebox{\plotpoint}}
\put(574.45,545.78){\usebox{\plotpoint}}
\put(586.60,562.61){\usebox{\plotpoint}}
\put(598.69,579.48){\usebox{\plotpoint}}
\multiput(610,596)(12.152,16.826){2}{\usebox{\plotpoint}}
\put(634.73,630.25){\usebox{\plotpoint}}
\put(646.89,647.07){\usebox{\plotpoint}}
\put(659.04,663.90){\usebox{\plotpoint}}
\put(671.53,680.47){\usebox{\plotpoint}}
\put(684.49,696.68){\usebox{\plotpoint}}
\put(697.58,712.79){\usebox{\plotpoint}}
\put(711.04,728.59){\usebox{\plotpoint}}
\put(724.63,744.27){\usebox{\plotpoint}}
\put(739.12,759.12){\usebox{\plotpoint}}
\put(754.44,773.11){\usebox{\plotpoint}}
\put(771.08,785.52){\usebox{\plotpoint}}
\put(788.47,796.83){\usebox{\plotpoint}}
\put(806.72,806.66){\usebox{\plotpoint}}
\put(826.45,812.95){\usebox{\plotpoint}}
\put(846.92,816.00){\usebox{\plotpoint}}
\put(867.68,816.00){\usebox{\plotpoint}}
\put(888.00,812.15){\usebox{\plotpoint}}
\put(907.84,806.05){\usebox{\plotpoint}}
\put(927.05,798.21){\usebox{\plotpoint}}
\put(945.89,789.51){\usebox{\plotpoint}}
\put(964.06,779.50){\usebox{\plotpoint}}
\put(982.03,769.14){\usebox{\plotpoint}}
\put(999.84,758.48){\usebox{\plotpoint}}
\put(1017.70,747.93){\usebox{\plotpoint}}
\put(1035.62,737.46){\usebox{\plotpoint}}
\put(1053.81,727.48){\usebox{\plotpoint}}
\put(1072.34,718.15){\usebox{\plotpoint}}
\put(1090.78,708.64){\usebox{\plotpoint}}
\put(1109.81,700.38){\usebox{\plotpoint}}
\put(1129.19,692.94){\usebox{\plotpoint}}
\put(1148.86,686.36){\usebox{\plotpoint}}
\put(1168.80,680.75){\usebox{\plotpoint}}
\put(1188.90,675.79){\usebox{\plotpoint}}
\put(1209.22,671.66){\usebox{\plotpoint}}
\put(1229.74,668.50){\usebox{\plotpoint}}
\put(1250.37,666.33){\usebox{\plotpoint}}
\put(1270.98,664.08){\usebox{\plotpoint}}
\put(1291.70,663.00){\usebox{\plotpoint}}
\put(1312.41,662.00){\usebox{\plotpoint}}
\put(1333.17,662.00){\usebox{\plotpoint}}
\put(1353.93,662.00){\usebox{\plotpoint}}
\put(1374.68,662.00){\usebox{\plotpoint}}
\put(1395.40,663.00){\usebox{\plotpoint}}
\put(1416.15,663.09){\usebox{\plotpoint}}
\put(1436.87,664.00){\usebox{\plotpoint}}
\put(1457.58,665.28){\usebox{\plotpoint}}
\put(1478.31,666.00){\usebox{\plotpoint}}
\put(1480,666){\usebox{\plotpoint}}
\sbox{\plotpoint}{\rule[-0.400pt]{0.800pt}{0.800pt}}%
\put(181,165){\usebox{\plotpoint}}
\multiput(181.00,166.39)(1.244,0.536){5}{\rule{1.933pt}{0.129pt}}
\multiput(181.00,163.34)(8.987,6.000){2}{\rule{0.967pt}{0.800pt}}
\multiput(194.00,172.39)(1.244,0.536){5}{\rule{1.933pt}{0.129pt}}
\multiput(194.00,169.34)(8.987,6.000){2}{\rule{0.967pt}{0.800pt}}
\multiput(207.00,178.39)(1.244,0.536){5}{\rule{1.933pt}{0.129pt}}
\multiput(207.00,175.34)(8.987,6.000){2}{\rule{0.967pt}{0.800pt}}
\multiput(220.00,184.40)(1.000,0.526){7}{\rule{1.686pt}{0.127pt}}
\multiput(220.00,181.34)(9.501,7.000){2}{\rule{0.843pt}{0.800pt}}
\multiput(233.00,191.40)(1.000,0.526){7}{\rule{1.686pt}{0.127pt}}
\multiput(233.00,188.34)(9.501,7.000){2}{\rule{0.843pt}{0.800pt}}
\multiput(246.00,198.40)(1.000,0.526){7}{\rule{1.686pt}{0.127pt}}
\multiput(246.00,195.34)(9.501,7.000){2}{\rule{0.843pt}{0.800pt}}
\multiput(259.00,205.40)(1.000,0.526){7}{\rule{1.686pt}{0.127pt}}
\multiput(259.00,202.34)(9.501,7.000){2}{\rule{0.843pt}{0.800pt}}
\multiput(272.00,212.40)(0.847,0.520){9}{\rule{1.500pt}{0.125pt}}
\multiput(272.00,209.34)(9.887,8.000){2}{\rule{0.750pt}{0.800pt}}
\multiput(285.00,220.40)(0.847,0.520){9}{\rule{1.500pt}{0.125pt}}
\multiput(285.00,217.34)(9.887,8.000){2}{\rule{0.750pt}{0.800pt}}
\multiput(298.00,228.40)(0.847,0.520){9}{\rule{1.500pt}{0.125pt}}
\multiput(298.00,225.34)(9.887,8.000){2}{\rule{0.750pt}{0.800pt}}
\multiput(311.00,236.40)(0.737,0.516){11}{\rule{1.356pt}{0.124pt}}
\multiput(311.00,233.34)(10.186,9.000){2}{\rule{0.678pt}{0.800pt}}
\multiput(324.00,245.40)(0.737,0.516){11}{\rule{1.356pt}{0.124pt}}
\multiput(324.00,242.34)(10.186,9.000){2}{\rule{0.678pt}{0.800pt}}
\multiput(337.00,254.40)(0.737,0.516){11}{\rule{1.356pt}{0.124pt}}
\multiput(337.00,251.34)(10.186,9.000){2}{\rule{0.678pt}{0.800pt}}
\multiput(350.00,263.40)(0.654,0.514){13}{\rule{1.240pt}{0.124pt}}
\multiput(350.00,260.34)(10.426,10.000){2}{\rule{0.620pt}{0.800pt}}
\multiput(363.00,273.40)(0.654,0.514){13}{\rule{1.240pt}{0.124pt}}
\multiput(363.00,270.34)(10.426,10.000){2}{\rule{0.620pt}{0.800pt}}
\multiput(376.00,283.40)(0.654,0.514){13}{\rule{1.240pt}{0.124pt}}
\multiput(376.00,280.34)(10.426,10.000){2}{\rule{0.620pt}{0.800pt}}
\multiput(389.00,293.40)(0.589,0.512){15}{\rule{1.145pt}{0.123pt}}
\multiput(389.00,290.34)(10.623,11.000){2}{\rule{0.573pt}{0.800pt}}
\multiput(402.00,304.40)(0.589,0.512){15}{\rule{1.145pt}{0.123pt}}
\multiput(402.00,301.34)(10.623,11.000){2}{\rule{0.573pt}{0.800pt}}
\multiput(415.00,315.41)(0.536,0.511){17}{\rule{1.067pt}{0.123pt}}
\multiput(415.00,312.34)(10.786,12.000){2}{\rule{0.533pt}{0.800pt}}
\multiput(428.00,327.41)(0.536,0.511){17}{\rule{1.067pt}{0.123pt}}
\multiput(428.00,324.34)(10.786,12.000){2}{\rule{0.533pt}{0.800pt}}
\multiput(441.00,339.41)(0.492,0.509){19}{\rule{1.000pt}{0.123pt}}
\multiput(441.00,336.34)(10.924,13.000){2}{\rule{0.500pt}{0.800pt}}
\multiput(454.00,352.41)(0.492,0.509){19}{\rule{1.000pt}{0.123pt}}
\multiput(454.00,349.34)(10.924,13.000){2}{\rule{0.500pt}{0.800pt}}
\multiput(467.00,365.41)(0.492,0.509){19}{\rule{1.000pt}{0.123pt}}
\multiput(467.00,362.34)(10.924,13.000){2}{\rule{0.500pt}{0.800pt}}
\multiput(481.41,377.00)(0.509,0.533){19}{\rule{0.123pt}{1.062pt}}
\multiput(478.34,377.00)(13.000,11.797){2}{\rule{0.800pt}{0.531pt}}
\multiput(494.41,391.00)(0.509,0.533){19}{\rule{0.123pt}{1.062pt}}
\multiput(491.34,391.00)(13.000,11.797){2}{\rule{0.800pt}{0.531pt}}
\multiput(507.41,405.00)(0.509,0.574){19}{\rule{0.123pt}{1.123pt}}
\multiput(504.34,405.00)(13.000,12.669){2}{\rule{0.800pt}{0.562pt}}
\multiput(520.41,420.00)(0.509,0.574){19}{\rule{0.123pt}{1.123pt}}
\multiput(517.34,420.00)(13.000,12.669){2}{\rule{0.800pt}{0.562pt}}
\multiput(533.41,435.00)(0.509,0.616){19}{\rule{0.123pt}{1.185pt}}
\multiput(530.34,435.00)(13.000,13.541){2}{\rule{0.800pt}{0.592pt}}
\multiput(546.41,451.00)(0.509,0.616){19}{\rule{0.123pt}{1.185pt}}
\multiput(543.34,451.00)(13.000,13.541){2}{\rule{0.800pt}{0.592pt}}
\multiput(559.41,467.00)(0.509,0.657){19}{\rule{0.123pt}{1.246pt}}
\multiput(556.34,467.00)(13.000,14.414){2}{\rule{0.800pt}{0.623pt}}
\multiput(572.41,484.00)(0.509,0.616){19}{\rule{0.123pt}{1.185pt}}
\multiput(569.34,484.00)(13.000,13.541){2}{\rule{0.800pt}{0.592pt}}
\multiput(585.41,500.00)(0.509,0.698){19}{\rule{0.123pt}{1.308pt}}
\multiput(582.34,500.00)(13.000,15.286){2}{\rule{0.800pt}{0.654pt}}
\multiput(598.41,518.00)(0.509,0.657){19}{\rule{0.123pt}{1.246pt}}
\multiput(595.34,518.00)(13.000,14.414){2}{\rule{0.800pt}{0.623pt}}
\multiput(611.41,535.00)(0.509,0.657){19}{\rule{0.123pt}{1.246pt}}
\multiput(608.34,535.00)(13.000,14.414){2}{\rule{0.800pt}{0.623pt}}
\multiput(624.41,552.00)(0.509,0.698){19}{\rule{0.123pt}{1.308pt}}
\multiput(621.34,552.00)(13.000,15.286){2}{\rule{0.800pt}{0.654pt}}
\multiput(637.41,570.00)(0.509,0.698){19}{\rule{0.123pt}{1.308pt}}
\multiput(634.34,570.00)(13.000,15.286){2}{\rule{0.800pt}{0.654pt}}
\multiput(650.41,588.00)(0.509,0.698){19}{\rule{0.123pt}{1.308pt}}
\multiput(647.34,588.00)(13.000,15.286){2}{\rule{0.800pt}{0.654pt}}
\multiput(663.41,606.00)(0.509,0.657){19}{\rule{0.123pt}{1.246pt}}
\multiput(660.34,606.00)(13.000,14.414){2}{\rule{0.800pt}{0.623pt}}
\multiput(676.41,623.00)(0.509,0.657){19}{\rule{0.123pt}{1.246pt}}
\multiput(673.34,623.00)(13.000,14.414){2}{\rule{0.800pt}{0.623pt}}
\multiput(689.41,640.00)(0.509,0.657){19}{\rule{0.123pt}{1.246pt}}
\multiput(686.34,640.00)(13.000,14.414){2}{\rule{0.800pt}{0.623pt}}
\multiput(702.41,657.00)(0.509,0.657){19}{\rule{0.123pt}{1.246pt}}
\multiput(699.34,657.00)(13.000,14.414){2}{\rule{0.800pt}{0.623pt}}
\multiput(715.41,674.00)(0.509,0.616){19}{\rule{0.123pt}{1.185pt}}
\multiput(712.34,674.00)(13.000,13.541){2}{\rule{0.800pt}{0.592pt}}
\multiput(728.41,690.00)(0.509,0.574){19}{\rule{0.123pt}{1.123pt}}
\multiput(725.34,690.00)(13.000,12.669){2}{\rule{0.800pt}{0.562pt}}
\multiput(741.41,705.00)(0.509,0.533){19}{\rule{0.123pt}{1.062pt}}
\multiput(738.34,705.00)(13.000,11.797){2}{\rule{0.800pt}{0.531pt}}
\multiput(754.41,719.00)(0.509,0.533){19}{\rule{0.123pt}{1.062pt}}
\multiput(751.34,719.00)(13.000,11.797){2}{\rule{0.800pt}{0.531pt}}
\multiput(766.00,734.41)(0.536,0.511){17}{\rule{1.067pt}{0.123pt}}
\multiput(766.00,731.34)(10.786,12.000){2}{\rule{0.533pt}{0.800pt}}
\multiput(779.00,746.40)(0.589,0.512){15}{\rule{1.145pt}{0.123pt}}
\multiput(779.00,743.34)(10.623,11.000){2}{\rule{0.573pt}{0.800pt}}
\multiput(792.00,757.40)(0.654,0.514){13}{\rule{1.240pt}{0.124pt}}
\multiput(792.00,754.34)(10.426,10.000){2}{\rule{0.620pt}{0.800pt}}
\multiput(805.00,767.40)(0.737,0.516){11}{\rule{1.356pt}{0.124pt}}
\multiput(805.00,764.34)(10.186,9.000){2}{\rule{0.678pt}{0.800pt}}
\multiput(818.00,776.40)(1.000,0.526){7}{\rule{1.686pt}{0.127pt}}
\multiput(818.00,773.34)(9.501,7.000){2}{\rule{0.843pt}{0.800pt}}
\multiput(831.00,783.38)(1.600,0.560){3}{\rule{2.120pt}{0.135pt}}
\multiput(831.00,780.34)(7.600,5.000){2}{\rule{1.060pt}{0.800pt}}
\multiput(843.00,788.38)(1.768,0.560){3}{\rule{2.280pt}{0.135pt}}
\multiput(843.00,785.34)(8.268,5.000){2}{\rule{1.140pt}{0.800pt}}
\put(856,791.84){\rule{3.132pt}{0.800pt}}
\multiput(856.00,790.34)(6.500,3.000){2}{\rule{1.566pt}{0.800pt}}
\put(869,793.84){\rule{3.132pt}{0.800pt}}
\multiput(869.00,793.34)(6.500,1.000){2}{\rule{1.566pt}{0.800pt}}
\put(882,794.84){\rule{3.132pt}{0.800pt}}
\multiput(882.00,794.34)(6.500,1.000){2}{\rule{1.566pt}{0.800pt}}
\put(895,794.84){\rule{3.132pt}{0.800pt}}
\multiput(895.00,795.34)(6.500,-1.000){2}{\rule{1.566pt}{0.800pt}}
\put(908,793.34){\rule{3.132pt}{0.800pt}}
\multiput(908.00,794.34)(6.500,-2.000){2}{\rule{1.566pt}{0.800pt}}
\put(921,790.84){\rule{3.132pt}{0.800pt}}
\multiput(921.00,792.34)(6.500,-3.000){2}{\rule{1.566pt}{0.800pt}}
\put(934,787.34){\rule{2.800pt}{0.800pt}}
\multiput(934.00,789.34)(7.188,-4.000){2}{\rule{1.400pt}{0.800pt}}
\multiput(947.00,785.06)(1.768,-0.560){3}{\rule{2.280pt}{0.135pt}}
\multiput(947.00,785.34)(8.268,-5.000){2}{\rule{1.140pt}{0.800pt}}
\multiput(960.00,780.06)(1.768,-0.560){3}{\rule{2.280pt}{0.135pt}}
\multiput(960.00,780.34)(8.268,-5.000){2}{\rule{1.140pt}{0.800pt}}
\multiput(973.00,775.07)(1.244,-0.536){5}{\rule{1.933pt}{0.129pt}}
\multiput(973.00,775.34)(8.987,-6.000){2}{\rule{0.967pt}{0.800pt}}
\multiput(986.00,769.07)(1.244,-0.536){5}{\rule{1.933pt}{0.129pt}}
\multiput(986.00,769.34)(8.987,-6.000){2}{\rule{0.967pt}{0.800pt}}
\multiput(999.00,763.08)(1.000,-0.526){7}{\rule{1.686pt}{0.127pt}}
\multiput(999.00,763.34)(9.501,-7.000){2}{\rule{0.843pt}{0.800pt}}
\multiput(1012.00,756.08)(1.000,-0.526){7}{\rule{1.686pt}{0.127pt}}
\multiput(1012.00,756.34)(9.501,-7.000){2}{\rule{0.843pt}{0.800pt}}
\multiput(1025.00,749.07)(1.244,-0.536){5}{\rule{1.933pt}{0.129pt}}
\multiput(1025.00,749.34)(8.987,-6.000){2}{\rule{0.967pt}{0.800pt}}
\multiput(1038.00,743.08)(1.000,-0.526){7}{\rule{1.686pt}{0.127pt}}
\multiput(1038.00,743.34)(9.501,-7.000){2}{\rule{0.843pt}{0.800pt}}
\multiput(1051.00,736.08)(1.000,-0.526){7}{\rule{1.686pt}{0.127pt}}
\multiput(1051.00,736.34)(9.501,-7.000){2}{\rule{0.843pt}{0.800pt}}
\multiput(1064.00,729.07)(1.244,-0.536){5}{\rule{1.933pt}{0.129pt}}
\multiput(1064.00,729.34)(8.987,-6.000){2}{\rule{0.967pt}{0.800pt}}
\multiput(1077.00,723.07)(1.244,-0.536){5}{\rule{1.933pt}{0.129pt}}
\multiput(1077.00,723.34)(8.987,-6.000){2}{\rule{0.967pt}{0.800pt}}
\multiput(1090.00,717.07)(1.244,-0.536){5}{\rule{1.933pt}{0.129pt}}
\multiput(1090.00,717.34)(8.987,-6.000){2}{\rule{0.967pt}{0.800pt}}
\multiput(1103.00,711.07)(1.244,-0.536){5}{\rule{1.933pt}{0.129pt}}
\multiput(1103.00,711.34)(8.987,-6.000){2}{\rule{0.967pt}{0.800pt}}
\multiput(1116.00,705.06)(1.768,-0.560){3}{\rule{2.280pt}{0.135pt}}
\multiput(1116.00,705.34)(8.268,-5.000){2}{\rule{1.140pt}{0.800pt}}
\multiput(1129.00,700.06)(1.768,-0.560){3}{\rule{2.280pt}{0.135pt}}
\multiput(1129.00,700.34)(8.268,-5.000){2}{\rule{1.140pt}{0.800pt}}
\put(1142,693.34){\rule{2.800pt}{0.800pt}}
\multiput(1142.00,695.34)(7.188,-4.000){2}{\rule{1.400pt}{0.800pt}}
\multiput(1155.00,691.06)(1.768,-0.560){3}{\rule{2.280pt}{0.135pt}}
\multiput(1155.00,691.34)(8.268,-5.000){2}{\rule{1.140pt}{0.800pt}}
\put(1168,684.84){\rule{3.132pt}{0.800pt}}
\multiput(1168.00,686.34)(6.500,-3.000){2}{\rule{1.566pt}{0.800pt}}
\put(1181,681.34){\rule{2.800pt}{0.800pt}}
\multiput(1181.00,683.34)(7.188,-4.000){2}{\rule{1.400pt}{0.800pt}}
\put(1194,678.34){\rule{3.132pt}{0.800pt}}
\multiput(1194.00,679.34)(6.500,-2.000){2}{\rule{1.566pt}{0.800pt}}
\put(1207,675.84){\rule{3.132pt}{0.800pt}}
\multiput(1207.00,677.34)(6.500,-3.000){2}{\rule{1.566pt}{0.800pt}}
\put(1220,673.34){\rule{3.132pt}{0.800pt}}
\multiput(1220.00,674.34)(6.500,-2.000){2}{\rule{1.566pt}{0.800pt}}
\put(1233,671.34){\rule{3.132pt}{0.800pt}}
\multiput(1233.00,672.34)(6.500,-2.000){2}{\rule{1.566pt}{0.800pt}}
\put(1246,669.34){\rule{3.132pt}{0.800pt}}
\multiput(1246.00,670.34)(6.500,-2.000){2}{\rule{1.566pt}{0.800pt}}
\put(1259,667.84){\rule{3.132pt}{0.800pt}}
\multiput(1259.00,668.34)(6.500,-1.000){2}{\rule{1.566pt}{0.800pt}}
\put(1272,666.34){\rule{3.132pt}{0.800pt}}
\multiput(1272.00,667.34)(6.500,-2.000){2}{\rule{1.566pt}{0.800pt}}
\put(1285,664.84){\rule{3.132pt}{0.800pt}}
\multiput(1285.00,665.34)(6.500,-1.000){2}{\rule{1.566pt}{0.800pt}}
\put(1311,663.84){\rule{3.132pt}{0.800pt}}
\multiput(1311.00,664.34)(6.500,-1.000){2}{\rule{1.566pt}{0.800pt}}
\put(1298.0,666.0){\rule[-0.400pt]{3.132pt}{0.800pt}}
\put(1337,662.84){\rule{3.132pt}{0.800pt}}
\multiput(1337.00,663.34)(6.500,-1.000){2}{\rule{1.566pt}{0.800pt}}
\put(1324.0,665.0){\rule[-0.400pt]{3.132pt}{0.800pt}}
\put(1402,662.84){\rule{3.132pt}{0.800pt}}
\multiput(1402.00,662.34)(6.500,1.000){2}{\rule{1.566pt}{0.800pt}}
\put(1350.0,664.0){\rule[-0.400pt]{12.527pt}{0.800pt}}
\put(1441,663.84){\rule{3.132pt}{0.800pt}}
\multiput(1441.00,663.34)(6.500,1.000){2}{\rule{1.566pt}{0.800pt}}
\put(1415.0,665.0){\rule[-0.400pt]{6.263pt}{0.800pt}}
\put(1467,664.84){\rule{3.132pt}{0.800pt}}
\multiput(1467.00,664.34)(6.500,1.000){2}{\rule{1.566pt}{0.800pt}}
\put(1454.0,666.0){\rule[-0.400pt]{3.132pt}{0.800pt}}
\put(1600,103){$ (+ \, 2m_t)$}
\end{picture}
\end{center}
\vspace{-7mm}

For the $\Upsilon(1S)$ the order $\als^5 \ln \als$ correction evaluated from Eq. (\ref{nnnlospectrum}) reads
\be
\delta E_{1,0,1} = {1730 \over 81 \pi} m_b \als^4(r) \als \ln (1/\als).
\ee
This correction has been considered in \cite{bn,yn}. 
In dependence of the (ultrasoft) scale at which $\als$ is calculated it may be 
as large as $80 - 100$  MeV. It is not clear, up to now, if the size of this correction
should be taken as a serious estimate of the complete order $\als^5$, or if it will largely cancel 
against the remaining $\als^5$ contributions (still unknown). It is also possible that this is a signal 
of the renormalon of order $\lQ^3 r^2$ affecting the static potential, which has proved
to be related to this kind of corrections in \cite{long}. 

Despite some remarkable progress achieved recently in increasing our know\-led\-ge of perturbative 
corrections either by resumming potentially large lo\-ga\-rithms \cite{RG} or by considering in 
the bottomonium system the effects due to the finite charm quark mass \cite{hoangmc}, 
the real challenge for heavy quarkonia remain non-perturbative contributions. The uncertainty related 
to them is usually be\-li\-ev\-ed to be of 100 MeV for the $\Upsilon(1S)$ and of several hundreds MeV 
for the $J/\psi$ and, therefore, it dominates over higher perturbative 
corrections, once a renormalon free mass definition has been used. The leading non-perturbative 
contributions to the spectrum can be also read from Eq. (\ref{nnnlospectrum}). 
For the general case $\lQ \siml m v^2$ they are encoded into the gluonic correlator 
$\langle {\bf E}^a (t) \phi(t,0)^{\rm adj}_{ab} {\bf E}^b (0) \rangle$, which may be expanded in 
terms of local condensates in the situation $\lQ \ll m v^2$. A discussion can be found in \cite{vil}.
It is not clear what situation applies to the physical systems of interest.  If the ground states 
of bottomonium and charmonium fall into the situation  $\lQ \siml m v^2$, which is likely, a study of these 
systems using Eq. (\ref{nnnlospectrum}) and one of the parameterization of the gluonic 
correlator suggested by sum-rule calculations \cite{eesum}, by different lattice 
simulations \cite{eelat1,eelat2} or by QCD vacuum models \cite{anto} is timely.

\begin{figure}[htb]
\makebox[0cm]{\phantom b}
\put(0,30){\epsfxsize=9truecm \epsfbox{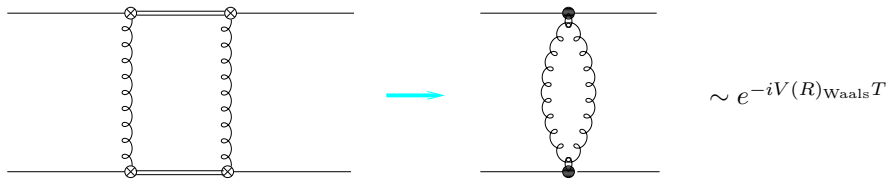}}
\put(270,62){$\sim e^{-i  V(R)_{\rm Waals} T}$}
\vspace{-9mm}
\caption{Quarkonium-quarkonium scattering and extraction of the van der Waals potential.}
\vspace{-3mm}
\label{figwal}
\end{figure}

\subsection{Quarkonium-quarkonium scattering}
The above EFT approach may be further pursued if the heavy quarkonium system interacts 
with other systems  so that scales smaller than the binding energy, $E^{\rm bind}$, are 
present. This is the situation that may happen in the scattering of heavy-quarkonium states 
if the energy of the hadron, $E^{\rm had}$, is much smaller than $E^{\rm bind}$ \cite{peskin79}. 
By integrating out from the scattering amplitude the higher energy scales, 
we may get a suitable definition of the quarkonium-quarkonium potential (see Fig. \ref{figwal}).
The situation is similar to the matching to pNRQCD discussed above. 
More specifically the quarkonium-quarkonium van der Waals  potential is given by 
\begin{eqnarray}
\hspace{-8mm} 
V(R)_{\rm Waals} &=& i  a^{ij} a^{lm} g^4 \int_{-\infty}^\infty \!\!\!\! d t \, 
\langle  E_j^a  E_i^a( t,R)  E_l^b E_m^b( 0,0)\rangle,  
\label{waals}
\end{eqnarray}
being $R$ the relative coordinate of the quarkonia. 
The coefficients $a^{ij}$ come from the matching to pNRQCD:
$$
a^{ij} = - V_A^2 {T_F\over N_c}  \langle \hbox{quarkonium state} \vert  
r_j {i \over E^{\rm bind} - h_o} r_i \vert \hbox{quarkonium state} \rangle.
$$
Notice that in this case the relevant non-perturbative operator is a four--chromoelectric-field correlator. 
Applications to the $J/\psi - J/\psi$ scattering have been recently discussed in \cite{fuj}.

\section{Long-range quarkonium}
\label{sec2}
For higher heavy-quarkonium states the Coulombic Bohr radius tends to become large 
and the perturbative matching, which led to pNRQCD in the above formulation, 
is no longer justified.\footnote{It is debated if for the charmonium ground state 
the scale of the momentum transfer may or not be integrated out perturbatively. 
Here, I note that, if this cannot be done, also the mentioned approach   
to the $J/\psi-J/\psi$ scattering needs to be revised.}
However, the success of traditional potential models seems to suggest that a NR description of these 
systems may still hold (for some reviews see \cite{rev0,rev1}). Therefore, one may still think 
to follow the same procedure discussed in the previous section for the perturbative case and integrate out the 
scale of the momentum transfer in order to get what would be pNRQCD in this situation. 
The result is a NR quantum-mechanical description of heavy quarkonium fully 
derived from (NR)QCD via a non-perturbative matching. I will assume that the matching between 
NRQCD and pNRQCD can be performed order by order in a $1/m$ expansion. 
While this can be justified within a perturbative framework, in a non-perturbative situation, one may 
question on its validity. For instance, a case where certain degrees of freedom cannot be integrated out 
in this way has been considered in \cite{DS}. This point surely deserves further studies. 
It is also relevant in order to establish the proper power counting of NRQCD, 
on which I will comment later on.

In order to identify the degrees of freedom of pNRQCD in the situation where the momentum 
transfer of the system is close to $\lQ$, I first consider the NRQCD Hamiltonian 
up to order $1/m^2$, $H = H^{(0)}+ \displaystyle\sum_{n=1,2} {H^{(n)} \over m^n}$ 
where $H^{(0)} = \displaystyle \int d^3{\bf x} {1\over 2}\left( {\bfPi}^a{\bfPi}^a +{\bf B}^a{\bf B}^a \right)$, 
$\bfPi^a$ is the canonical momentum conjugated to the gauge field ${\bf A}^a$, 
and the terms $H^{(1)}$ and $H^{(2)}$ may be read off from the NRQCD Lagrangian \cite{m2}. 
In the static limit the one-quark--one-antiquark sector of the Fock space may be spanned by
$\vert \underbar{n}; {\bf x}_1 ,{\bf x}_2  \rangle^{(0)}$ $=$ 
$\psi^{\dagger}({\bf x}_1) \chi_c^{\dagger} ({\bf x}_2) |n;{\bf x}_1 ,{\bf x}_2\rangle^{(0)}$, 
where $|\underbar{n}; {\bf x}_1 ,{\bf x}_2\rangle^{(0)} $ is a gauge-invariant
eigenstate (up to a phase) of $ H^{(0)}$ with energy $E_{n}^{(0)}({\bf x}_1 ,{\bf x}_2)$, 
and  $\chi_c ({\bf x})=i\sigma^2 \chi^{*} ({\bf x})$.  
$|n;{\bf x}_1 ,{\bf x}_2\rangle^{(0)}$ encodes the gluonic content of the
state, i.e.  it is annihilated by $\chi_c({\bf x})$ and $\psi ({\bf x})$ ($\forall {\bf x}$). 
The positions ${\bf x}_1$ and ${\bf x}_2$ of the quark and antiquark respectively 
are good quantum numbers for the static solution $|\underbar{n};{\bf x}_1 ,{\bf x}_2 \rangle^{(0)}$ 
(but will be used also to label the eigenstates of $H$); $n$ generically denotes the remaining 
quantum numbers, which are classified by the irreducible representations of the symmetry 
group $D_{\infty h}$ (substituting the parity generator by CP). I also choose 
$|\underbar{n};{\bf x}_1,{\bf x}_2 \rangle^{(0)}$ to be invariant under time inversion.
The ground-state energy $E_0^{(0)}({\bf x}_1,{\bf x}_2)$ can be associated (in some specific situation) 
to the static potential of the heavy quarkonium. The remaining energies 
$E_n^{(0)}({\bf x}_1,{\bf x}_2)$, $n\not=0$, are usually associated 
to the potentials describing heavy hybrids or heavy quarkonium (or other heavy hybrids) plus glueballs. 
They can be computed on the lattice (see, for instance, \cite{michael,nora99}). 
Beyond the static limit, but still working order by order in $1/m$, the eigenvalues 
$E_n({\bf x}_1 ,{\bf x}_2; {\bf p})$ of the Hamiltonian $H$, up to $O(1/m^2)$, are given by 
\bea
&&\!\!\!\!\!\!
E_n({\bf x}_1,{\bf x}_2; {\bf p})
\prod_{j=1}^2 \delta^{(3)} ({\bf x}_j^\prime -{\bf x}_j) =  
E_n^{(0)}({\bf x}_1,{\bf x}_2) 
\prod_{j=1}^2 \delta^{(3)} ({\bf x}_j^\prime -{\bf x}_j) \nn\\
&&\!\!\!\!\!\! 
+\, ^{(0)}\langle \underbar{n}; {\bf x}_1 ,{\bf x}_2 \vert
\sum_{j=1,2} {H^{(j)} \over m^j} 
\vert \underbar{n}; {\bf x}_1^\prime ,{\bf x}_2^\prime \rangle^{(0)} 
\nn \\
&&\!\!\!\!\!\!
- \, \sum_{k\neq n} \int d^3y_1 \, d^3y_2 \,
^{(0)}\langle \underbar{n}; {\bf x}_1 ,{\bf x}_2 \vert {H^{(1)} \over m} 
\vert \underbar{k}; {\bf y}_1 ,{\bf y}_2\rangle^{(0)}\,
^{(0)}\langle \underbar{k}; {\bf y}_1,{\bf y}_2 \vert {H^{(1)} \over m} 
\vert \underbar{n}; {\bf x}_1^\prime,{\bf x}_2^\prime \rangle^{(0)} 
\nn\\ &&\qquad \times {1\over 2}
\left( {1\over E_k^{(0)}({\bf y}_1,{\bf y}_2) - E_n^{(0)}({\bf x}_1^\prime,{\bf x}_2^\prime)} 
\right.
\left. + {1\over E_k^{(0)}({\bf y}_1,{\bf y}_2) - E_n^{(0)}({\bf x}_1,{\bf x}_2)} \right).
\label{En2}
\eea  
$E_0$ corresponds to the quantum-mechanical Hamiltonian of the heavy quarkonium (in some specific 
situation). The other energies $E_n$, for $n \!> \!0$, are related to the quantum-mechanical Hamiltonian 
of  higher gluonic excitations between heavy quarks. Explicit expressions for the energies $E_n$, 
obtained from the above formula, can be found in \cite{m1,m2}. 

Let me now assume that, because of a mass gap in QCD, the energy splitting between the ground state 
and the first gluonic excitation is larger than $mv^2$ (see also the data reported by G. Bali at this conference),
and, because of chiral symmetry breaking of QCD, Goldstone bosons (pions/kaons) appear. 
Hence, in this situation, the states with ultrasoft energies (i.e. the degrees of freedom of pNRQCD) 
would be the ultrasoft excitations about the static ground state, which we call the singlet, 
plus the Goldstone bosons. If one switches off the light fermions (pure gluodynamics), 
only the singlet survives and pNRQCD reduces to a pure two-particle NR quantum-mechanical system. 
Therefore, under the assumption of the validity of the $1/m$ expansion (in the matching) 
and of the existence of a mass gap between the singlet and the other gluonic excitations 
between heavy quarks, we obtain the typical situation described by potential models.
More specifically, in terms of the NRQCD states discussed above, this means that only  
$|\underbar{0}; {\bf x}_1 ,{\bf x}_2\rangle^{(0)}$ is kept as an explicit degree of freedom,  
whereas $|\underbar{n}; {\bf x}_1 ,{\bf x}_2\rangle^{(0)}$ with  $n\not=0$ are integrated 
out. $|\underbar{0}; {\bf x}_1 ,{\bf x}_2\rangle^{(0)}$ provides the only dynamical degree of freedom 
of the theory. It is described by means of a bilinear colour singlet field, $S({\bf x}_1,{\bf x}_2,t)$,  
with the same quantum numbers and transformation properties under symmetries. 
In the above situation, the Lagrangian of pNRQCD reads 
\be
{\cal L}_{\rm pNRQCD} = S^\dagger \bigg( i\partial_0 - h_s({\bf x}_1,{\bf x}_2,{\bf p})\bigg) S, 
\label{pnrqcdl}
\ee
where $h_s = \displaystyle{{\bf p}^2\over m} - \displaystyle{{\bf p}^4\over 4m^3}
+ V^{(0)} +  \displaystyle{V^{(1)} \over m} + \displaystyle{V^{(2)} \over m^2}$ 
is the Hamiltonian of the singlet and may be identified through the matching condition 
\be 
E_0({\bf x}_1,{\bf x}_2, {\bf p}) = h_s({\bf x}_1,{\bf x}_2,{\bf p}).
\label{singlet}
\ee
I note that, if other ultrasoft degrees of freedom, apart from the singlet, exist, they may be added 
systematically to the above Lagrangian in an analogous way as done in the perturbative 
situation (see Eq. (\ref{pnrqcdph})). For what concerns the effects on the computation 
of the potentials, since we are integrating over all the states, in the situation where some of them, 
different from the singlet, are ultrasoft, we would just need to subtract their contribution later on. 

In \cite{m1,m2} the matching of NRQCD to pNRQCD has been performed up to order $1/m^2$
and the above potentials have been obtained explicitly in terms of Wilson loops \cite{wilson}.
For the static potential the result reads 
\be 
V^{(0)} = \lim_{T\to\infty}{i\over T} \ln \langle W_\Box \rangle, 
\label{v0}
\ee
where $W_\Box$ is a rectangular Wilson loop of dimension $r\times T$. The $1/m$ and $1/m^2$ potentials 
may be read off from \cite{m1,m2} after the identifications (cf. Eq. (\ref{VVs2})):
\bea
& & V^{(1)} = 2 V^{(1,0)}, \qquad 
V_{{\bf p}^2}^{(2)} = V_{{\bf p}^2}^{(2,0)} + {V_{{\bf p}^2}^{(1,1)}\over 2}, \qquad  
V_{{\bf L}^2}^{(2)} = 2  V_{{\bf L}^2}^{(2,0)} + V_{{\bf L}^2}^{(1,1)}, \nn\\ 
& & V_r^{(2)} = 2 V_r^{(2,0)} + V_r^{(1,1)}, \qquad  
V_{LS}^{(2)} = V_{LS}^{(2,0)} + V_{L_1S_2}^{(1,1)}, \nn\\
& & V_{S^2}^{(2)} = V_{S^2}^{(1,1)}, \qquad  
V_{{S}_{12}}^{(2)} = V_{{S}_{12}}^{(1,1)}. 
\label{Em2}
\eea
Having expressed the non-perturbative dynamics of the heavy-quark potentials 
in terms of Wilson loops is extremely convenient, for these quantities 
may be calculated directly in lattice simulations \cite{latpot}.  
Moreover, these operators can be also calculated in QCD vacuum models \cite{mod,flux},  
providing a way to check, directly on the phenomenology, assumptions on the 
structure of the QCD vacuum. In particular, it would be of interest to see what the vortices picture of the 
QCD vacuum, which has received so much attention in recent years \cite{rein}, 
may predict on these correlators. Finally, I would like to stress that the obtained expressions 
for the potentials are also correct perturbatively at any order in $\als$.

\subsection{The NRQCD power counting}
An important issue, once the EFT Lagrangian has been calculated through the matching 
procedure, is to establish its power counting in order to calculate physical observables.\footnote{
It should be stressed that the way the matching procedure is organized, e.g. as an $1/m$ 
expansion, may be different from the power counting of the EFT.} 
Establishing the power counting of pNRQCD in the non-perturbative regime is, however, not 
only important by itself. It may also serve to establish the non-perturbative power counting 
of NRQCD, which is an important source of information on the spectrum of excited quarkonium 
states (but also heavy-light mesons) \cite{sara}. The power counting usually adopted there and discussed, 
for instance, in \cite{pc} is inherited from the perturbative regime. However, there is no certainty that 
this is the suitable one for calculating higher quarkonium states, since in the non-perturbative 
regime different power countings are, in principle, possible.
The above formulation of pNRQCD has translated the problem of the NRQCD power counting 
to obtaining the power counting of the different potentials. This is expected to be of some 
advantage: $1)$ because the power counting of pNRQCD is simpler and quantum-mechanical arguments, 
like the virial theorem, may be more properly formulated in this context; $2)$ because all the potentials 
are expressed in terms of Wilson loops, for which there are or there will be direct lattice measurements.  

As an example of power counting in pNRQCD we can assume that the potentials scale with $mv$.  
By definition the kinetic energy counts as $mv^2$. $V^{(0)}$ would count as $mv$, if   
the virial theorem would not constrain it to count also as $mv^2$. In the perturbative case 
this extra $v$ suppression comes from the factor $\als \sim v$ in the potential (see Eq. (\ref{newpot0})). 
From our general assumption $V^{(1)}/m$ scales like $mv^2$. Therefore, it could be in principle as
large as $V^{(0)}$. This makes a (still to do) lattice evaluation of this potential quite interesting.
Perturbatively, due to the factor $\als^2$ (see Eq. (\ref{Vs1})), it is $O(mv^4)$.
For what concerns the $1/m^2$ potentials, they are in this scheme of order $mv^3$. 
However, also here several constraints apply. Terms involving $V^{(0)}$ are suppressed 
by an extra factor $v$, due to the virial theorem. General symmetry relations \cite{spin,BMP},
also further constrain the power counting.  Finally, some of the potentials are $O(\als)$ 
suppressed in the matching coefficients inherited from NRQCD.

\section*{Acknowledgments}
I thank Nora Brambilla for reading the manuscript and comments.  
I thank the organizers for the invitation, the Alexander von Humboldt foundation for support 
and the CFIF at the Instituto Superior Tecnico of Lisbon for hospitality during the last stage 
of this work.


\section*{References}
\begin{thebibliography}{999}
\bibitem{NRQED} W. E. Caswell and G. P. Lepage, \PLB {\bf 167}, 437 (1986).
\bibitem{topnnlo} A. Hoang at al., {\em Eur. Phys. J. direct} C{\bf 3}, 1 (2000).
\bibitem{NRQCD} B. A. Thacker and G. P. Lepage, \PRD {\bf 43}, 196 (1991).
\bibitem{wilson} K. G. Wilson, \PRD {\bf 10}, 2445 (1974). 
\bibitem{Brown} L. Susskind, Les Houches lectures (1976); 
 W. Fischler, \NPB {\bf 129}, 157 (1977);
 L. S. Brown and W. I. Weisberger, \PRD {\bf 20}, 3239 (1979). 
\bibitem{spin}  E. Eichten and F. L. Feinberg, \PRD {\bf 23}, 2724 (1981);
 M. E. Peskin, in Proceeding of the 11th SLAC Institute, SLAC Report No. 207, 151, 
 edited by P. Mc Donough (1983); D. Gromes, \ZPC {\bf 26}, 401 (1984); 
 in {\em Spectroscopy of light and heavy quarks}, ed. by  
 U. Gastaldi, R. Klapisch and F. Close, (Plenum Press, New York, 1989).
\bibitem{BMP}   A. Barchielli et al., \NPB {\bf 296}, 625 (1988); 
 (E) {\bf 303}, 752 (1988); 
 A. Barchielli et al., \NCA {\bf 103}, 59 (1990);
 N. Brambilla et al., \PRD {\bf 50}, 5878 (1994).
\bibitem{VL} M. B. Voloshin, \NPB {\bf 154}, 365 (1979); 
 {\em Sov. J. Nucl. Phys.} {\bf 36}, 143 (1982); 
 H. Leutwyler, \PLB {\bf 98}, 447 (1981).
\bibitem{gro82} D. Gromes, \PLB {\bf 115}, 482 (1982).
\bibitem{Balitsky} I. I. Balitsky, \NPB  {\bf 254}, 166 (1985). 
\bibitem{chen} Y. Chen et al., \PRD {\bf 52}, 264 (1995);
 N. Brambilla and A. Vairo, \NPB (Proc. Suppl.) {\bf 74}, 201 (1999). 
\bibitem{long} N. Brambilla et al., \PRD {\bf 60}, 091502 (1999); 
 \NPB {\bf 566}, 275 (2000).
\bibitem{m1} N. Brambilla et al., hep-ph/0002250. 
\bibitem{m2} A. Pineda and A. Vairo, hep-ph/0009145.
\bibitem{gois} S. Godfrey and N. Isgur, \PRD {\bf 32}, 189 (1985).
\bibitem{pNRQCD} A. Pineda and J. Soto, \NPB (Proc. Suppl.) {\bf 64}, 428 (1998).
\bibitem{1loop} A. Billoire, \PLB {\bf 92}, 343  (1980).
\bibitem{twoloop}  Y. Schr\"oder, \PLB {\bf 447}, 321 (1999);
 M. Peter, \PRL {\bf 78}, 602  (1997).
\bibitem{Duncan} A. Duncan, \PRD {\bf 13}, 2866 (1976). 
\bibitem{nnnlo} N. Brambilla et al., \PLB {\bf 470}, 215 (1999).
\bibitem{Gupta} S. N. Gupta et al., \PRD {\bf 26}, 3305 (1982); 
 S. Titard and F. J. Yndur{\'a}in, \PRD {\bf 49}, 6007 (1994). 
\bibitem{KP}  B. Kniehl and A. Penin, \NPB {\bf 563}, 200 (1999).
\bibitem{KP2} B. Kniehl and A. Penin, \NPB {\bf 577}, 197 (2000).
\bibitem{eeper} M. Eidem\"uller and M. Jamin, \PLB {\bf 416}, 415 (1998).
\bibitem{bn} M. Beneke, hep-ph/9911490.
\bibitem{yn} F. Yndur\'ain, hep-ph/0007333; hep-ph/0002237.
\bibitem{RG}  M. Luke et al., \PRD {\bf 61}, 074025 (2000);
 A. Manohar and I. Stewart, \PRD {\bf 62}, 014033 (2000); hep-ph/0003107;
 A. Manohar et al., \PLB {\bf 486}, 400 (2000); J. Soto and A. Pineda, hep-ph/0007197.
\bibitem{hoangmc} A. Hoang, hep-ph/0008102; D. Eiras and J. Soto, hep-ph/0005066.
\bibitem{vil} N. Brambilla and A. Vairo, hep-ph/0004192.
\bibitem{eesum} H. G. Dosch et al., \PLB {\bf 452}, 379 (1999).
\bibitem{eelat1} M. D'Elia et al., \PLB  {\bf 408}, 315 (1997).
\bibitem{eelat2} G. S. Bali et al., \PLB {\bf 421}, 265 (1998).
\bibitem{anto} M. Baker et al., \PRD {\bf 58}, 034010 (1998);
 D. Antonov, {\em JHEP} {\bf 07}, 055 (2000).
\bibitem{peskin79} M. E. Peskin, \NPB {\bf 156}, 365 (1979); 
 G. Bhanot and M. E. Peskin, \NPB {\bf 156}, 391 (1979); 
 see also B. R. Holstein contribution at INT 00-2 (2000).
\bibitem{fuj} H. Fujii and D. Kharzeev, \PRD {\bf 60}, 114039 (1999).
\bibitem{rev0} N. Brambilla and A. Vairo, hep-ph/9904330.
\bibitem{rev1} F. J. Yndur\'ain, hep-ph/9910399; \NPB (Proc. Suppl.) {\bf 64}, 433 (1998).
\bibitem{DS} D. Eiras and J. Soto, \PRD {\bf 61}, 114027 (2000).
\bibitem{michael} C. Michael, hep-ph/9809211 and at this conference; 
 K. J. Juge et al., hep-lat/9809015.
\bibitem{nora99} N. Brambilla, \NPB (Proc. Suppl.) {\bf 86}, 389 (2000).
\bibitem{latpot} G. S. Bali et al., \PRD {\bf 56}, 2566 (1997).
\bibitem{mod} N. Brambilla and A. Vairo, \PRD {\bf 55}, 3974 (1997); 
 M. Baker et al., \PLB {\bf 389}, 577 (1996);
 L. Fulcher, \PRD {\bf 62}, 094505 (2000) and at this conference.
\bibitem{flux} N. Brambilla, hep-ph/9809263. 
\bibitem{rein} H. Reinhardt at this conference; S. Olejnik at this conference.
\bibitem{sara} H. Shanahan at this conference; S. Collins at this conference. 
\bibitem{pc} G. P. Lepage et al., \PRD {\bf 46}, 4052 (1992).
\end{thebibliography}
\end{document}